\documentclass[prl,twocolumn,amsmath,amsfonts,superscriptaddress]{revtex4}

\usepackage{graphicx,float}
\usepackage{braket}
\usepackage{xcolor}
\usepackage{siunitx}

\DeclareMathOperator{\diag}{diag}

\begin{document}
    \title{Embedding entanglement generation within a measurement-feedback coherent Ising machine}
    \author{Ryotatsu~Yanagimoto}
    \email[Email: ]{ryotatsu@stanford.edu}
    \affiliation{E.\,L.\ Ginzton Laboratory, Stanford University, Stanford, California 94305, USA}
    \author{Peter~L.~McMahon}
    \affiliation{E.\,L.\ Ginzton Laboratory, Stanford University, Stanford, California 94305, USA}
    \affiliation{National Institute of Informatics, 2-1-2 Hitotsubashi, Chiyoda-ku, Tokyo 101-8430, Japan}
    \affiliation{School of Applied and Engineering Physics, Cornell University, Ithaca, New York 14853, USA}
    \author{Edwin~Ng}
    \affiliation{E.\,L.\ Ginzton Laboratory, Stanford University, Stanford, California 94305, USA}
    \author{Tatsuhiro~Onodera}
    \affiliation{E.\,L.\ Ginzton Laboratory, Stanford University, Stanford, California 94305, USA}
    \author{Hideo~Mabuchi}
    \email[Email: ]{hmabuchi@stanford.edu}
    \affiliation{E.\,L.\ Ginzton Laboratory, Stanford University, Stanford, California 94305, USA}

\date{\today}
\begin{abstract}
We present a new scheme to efficiently establish entanglement between optical modes in a time-multiplexed coherent Ising machine (CIM) by means of nonlocal measurement and feedback. We numerically simulate and evaluate the generation of steady-state entanglement in a system with nearest-neighbor interactions on a 1D ring, and we compare the results to those of a conventional CIM with all-optical interactions mediated by delay lines. We show that the delay-line architecture
has a fundamental limit on accessible entanglement that our measurement-based scheme can substantially surpass. These results motivate the study of non-classical correlations in CIMs with levels of entanglement previously considered to be experimentally impractical.
\end{abstract}

\maketitle

The Ising problem is a combinatorial optimization problem for which efficient mappings from many other important optimization problems are known~\cite{Lucas2014}. Formally, the Ising problem is to find the spin configurations $\sigma_k=\pm1$ which minimize the energy of the Ising Hamiltonian $-\sum_{k\ne \ell}J_{k\ell}\sigma_k\sigma_\ell$, given a coupling matrix $J_{k\ell}$. While finding optimal solutions to the Ising problem is presently intractable on conventional computers~\cite{Barahona1982}, the physical nature of the problem makes it especially amenable to straightforward physical realization. As such, many recent efforts have focused on the development of special-purpose hardware which can directly encode Ising spins and their couplings natively~\cite{Kim2010,Yamaoka2016,Johnson2011}.

In the optical domain, the coherent Ising machine (CIM) encodes spins using the two stable states of a degenerate optical parametric oscillator above threshold~\cite{Yamamoto2017,Wang2013}.  CIMs based on optical pulses have been experimentally realized at large scale~\cite{Inagaki2016,McMahon2016} using a measurement-feedback architecture, with favorable performance compared to the D-Wave quantum annealer~\cite{Hamerly2018}. However, as this architecture utilizes only local measurements and classical feedback (i.e., LOCC), it does not produce entanglement and the system is classically simulable~\cite{Vidal2003}. In order to study the potential for coherent or quantum advantages in CIMs (if any exist), it is important to better understand the nature of non-classical correlations which can be efficiently realized in CIMs in practice.

In this research, we present a scheme to generate interactions between the optical pulses of a time-multiplexed CIM by means of nonlocal measurement and feedback~\cite{Furusawa1998,Bouwmeester1997,Zukowski1993,Takei2005,Pan1998,Jia2004}, and we introduce an EPR-type measure to evaluate the resulting entanglement. For simplicity, we consider Ising networks with nearest-neighbor couplings on a 1D ring without frustration. We compare our scheme against an all-optical CIM architecture based on optical delay lines~\cite{Marandi2014,Takata2016}, and we show that, among other benefits, our measurement-based scheme can significantly surpass a fundamental limit on the steady-state entanglement accessible with the delay-line architecture.

We model a time-multiplexed CIM as an $N$-mode optical system, where the $k$th mode is associated with the bosonic annihilation operator $\hat{a}_k$. Then, we have the position-like operators $\hat{x}_k=(\hat{a}_k+\hat{a}_k^\dagger)/\sqrt{2}$ and the momentum-like operators $\hat{p}_k=(\hat{a}_k-\hat{a}_k^\dagger)/\sqrt{2}\mathrm{i}$, which can be summarized as a vector $\boldsymbol{\hat{R}}=(\hat{x}_1,\hat{p}_1,\dots,\hat{x}_N,\hat{p}_N)^\mathrm{T}$. Under the Gaussian approximation~\cite{Olivares2012}, all the information about the state $\hat \rho$ of the CIM is contained in the first-moments vector $\langle \boldsymbol{\hat{R}}\rangle=\mathrm{Tr}(\hat\rho\boldsymbol{\hat{R}})$ and the covariance matrix $\sigma$ with elements $\sigma_{k,\ell}=\frac{1}{2}\langle\{\hat{R}_k,\hat{R}_\ell\}\rangle-\langle\hat{R}_k\rangle\langle\hat{R}_\ell\rangle$, where $\{\hat{A},\hat{B}\}=\hat{A}\hat{B}+\hat{B}\hat{A}$. Since $\langle \boldsymbol{\hat{R}}\rangle$ can be shifted via only local displacements, the covariance matrix suffices to fully capture the non-classical correlations in the system. Therefore, we assume $\langle \boldsymbol{\hat{R}}\rangle=0$ and focus on the covariance matrix unless otherwise specified. Because the architectures we consider in this research do not introduce correlations between the $x$- and $p$-quadratures, we can assume $\langle\{\hat{x}_k,\hat{p}_\ell\}\rangle=0$, and it is convenient to introduce the notation $\sigma^\alpha_{k,\ell}=\frac{1}{2}\langle\{\hat{\alpha}_k,\hat{\alpha}_\ell\}\rangle-\langle\hat{\alpha}_k\rangle\langle\hat{\alpha}_\ell\rangle$ for $\alpha\in\{\text{x},\text{p}\}$ to denote the $x$-$x$ and $p$-$p$ correlations in $\sigma$.

\begin{figure}[ht]
    \centering
    \includegraphics[width=0.4\textwidth]{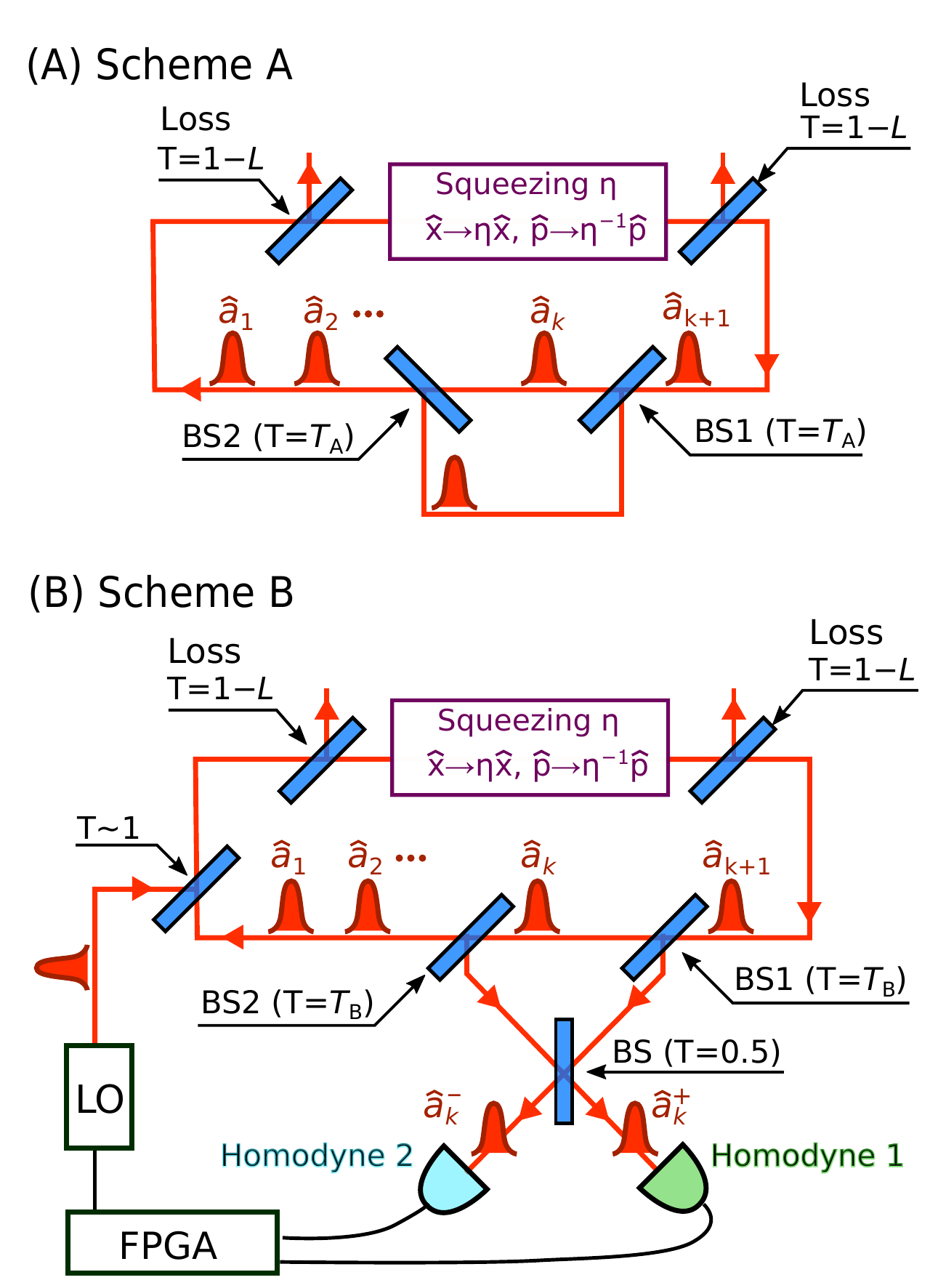}
    \caption{Schemes for generating interactions between pulses in a time-multiplexed CIM. (A) Scheme A uses an optical delay line adjunct to the CIM cavity. (B) Scheme B uses nonlocal measurement and feedback (see main text for details). To cancel the backaction on the mean displacement, feedback takes place through a beamsplitter with negligible reflectivity. Note that each homodyne setup involves mixing the input with an LO on a 50/50 BS followed by balanced detection. (BS: beamsplitter; T: (power) transmissivity; FPGA: Field-programmable gate array (for processing measurement results); LO: local oscillator)}
    \label{fig:schematics}
\end{figure}

One established way to generate interactions between pulses of a time-multiplexed CIM is to use an optical delay line as shown in Fig.~\ref{fig:schematics}(A). In this scheme (scheme A), a fraction of each mode is sent by a beamsplitter BS1 with transmissivity $T_\text{A}$ into a delay line to be mixed with the following mode through a beamsplitter BS2 with the same transmissivity $T_\text{A}$. The parity of the coupling $s_k=\pm1$ ($s_k=+1$ for ferromagnetic coupling and $-1$ for antiferromagnetic) between $\hat{x}_k$ and $\hat{x}_{k+1}$ is implemented by controlling the phase shift of the delay line.

In both scheme A and scheme B described below, we assume that every round trip, each pulse experiences squeezing (due to parametric gain) described by the mapping $\hat{x}_k\mapsto\eta\hat{x}_k$ and $\hat{p}_x\mapsto\eta^{-1}\hat{p}_x$ for some $\eta>1$, as well as some (intrinsic) loss $L_\mathrm{tot}$, modelled as two equal beamsplitters with reflectivities $L=1-\sqrt{1-L_\mathrm{tot}}$ just before and after the squeezing (e.g., representing the interface losses of a nonlinear crystal).

On the other hand, in the scheme we introduce in this research (scheme B), nearest-neighbor interactions are indirectly mediated through measurement and feedback as shown in Fig.~\ref{fig:schematics}(B). In this scheme, the interaction between target modes $k$ and $(k+1)$ is established in the following way. First, modes $k$ and $(k+1)$ impinge on two beamsplitters BS1 and BS2 with equal transmissivities $T_\mathrm{B}$, respectively. This generates two modes which remain in the cavity and two modes which are sent out of the cavity, which we denote respectively as $\hat{a}_\ell^\mathrm{i}=\sqrt{T_\mathrm{B}}\hat{a}_\ell+\sqrt{1-T_\mathrm{B}}\hat{a}^\mathrm{v}_\ell$ ($\ell=k,k+1$) and $\hat{a}_\ell^\mathrm{o}=\sqrt{1-T_\mathrm{B}}\hat{a}_\ell-\sqrt{T_\mathrm{B}}\hat{a}_\ell^\mathrm{v}$ ($\ell=k,k+1$); here, $\hat{a}_\ell^\mathrm{v}$ ($\ell=k,k+1$) represent the two vacuum modes which also impinge on BS1 and BS2, respectively.

Next, the two modes $\hat a_k^\mathrm{o}$ and $\hat a_{k+1}^\mathrm{o}$ sent out of the cavity are mixed by a 50/50 beamsplitter to produce the modes $\hat{a}_k^{\pm}=(\hat{a}^\mathrm{o}_k\mp\hat{a}^\mathrm{o}_{k+1})/\sqrt{2}$. In the limit of strong anti-squeezing (i.e., $\eta \gg 1$), $\hat{x}_\ell$ dominates over $\hat{x}_\ell^\mathrm{v}$, so up to constant factors, $\hat{x}_\ell^\mathrm{o}$ is equivalent to $\hat{x}_\ell$ which in turn is equivalent to $\hat{x}_\ell^\mathrm{i}$. Therefore, the measurement of the $x$-quadrature of $\hat a_k^{\pm}$ is equivalent to the measurement of $\hat{x}_{k+1}^\mathrm{i}\mp\hat{x}_{k}^\mathrm{i}$. We can then apply a feedback protocol based on this result to cancel the backaction on the first moments vector and thus enforce the relationship $\hat{x}_{k+1}^\mathrm{i}\mp\hat{x}_{k}^\mathrm{i}=0$~\cite{Takei2005}. Thus, in order to realize the interaction $s_k=\pm1$, we can simply 
measure the $x$-quadrature of $\hat{a}_k^{\pm}$ and the $p$-quadrature of $\hat{a}_k^{\mp}.$

Although our system only interacts nearest-neighbor pulses, we observe that the generated states show characteristic correlations between distant modes that exponentially decay with separation. That is, the covariance matrix at steady state approximately has the form
\begin{small}
\begin{subequations} \label{model}
\begin{align}
    \sigma_{k,\ell}^\mathrm{x}&=D^\mathrm{x}\delta_{k,\ell}+d^\mathrm{x}S_{k,\ell}\left[(\gamma^\mathrm{x})^{|k-\ell|}+S^*(\gamma^\mathrm{x})^{N-|k-\ell|}\right]\\
    \sigma_{k,\ell}^\mathrm{p}&=D^\mathrm{p}\delta_{k,\ell}-d^\mathrm{p}S_{k,\ell}\left[(\gamma^\mathrm{p})^{|k-\ell|}+S^*(\gamma^\mathrm{p})^{N-|k-\ell|}\right],
\end{align}
\end{subequations}
\end{small}where
$D^\alpha$, $d^\alpha$, and $\gamma^\alpha$ (for $\alpha\in\{\mathrm{x},\mathrm{p}\}$) are positive constants. Here, $S_{k,\ell}=\prod_{i=\min(k,\ell)}^{\max{(k,\ell)}-1}s_i$ is the coupling between $\hat{x}_k$ and $\hat{x}_\ell$ via all the modes in the middle (in this notation, $s_N$ represents the coupling between $\hat{x}_N$ and $\hat{x}_1$), while $S^*=s_1s_2\dots s_N$ represents the coupling of a mode to itself through all the rest of the modes. Therefore, $S^*=-1$ indicates the existence of frustration.

\begin{figure*}
\centering
\includegraphics[width=0.93\textwidth]{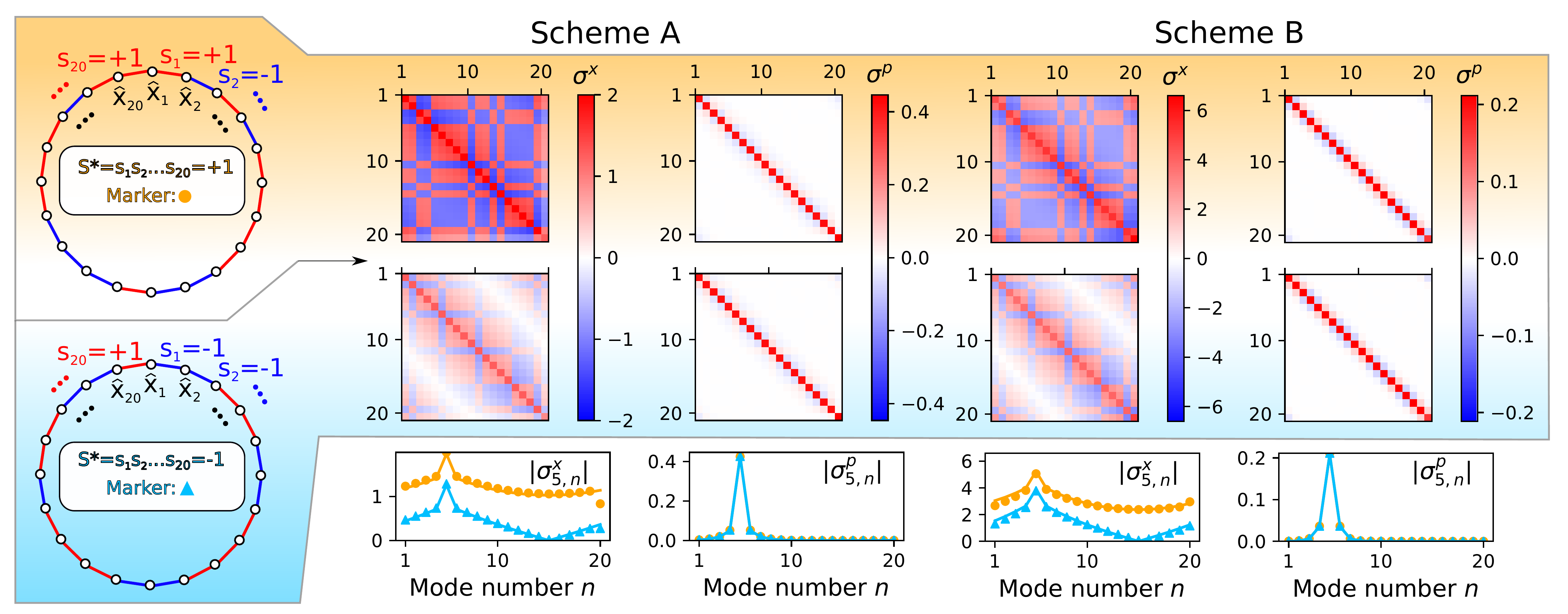}
\caption{Numerically-simulated steady-state covariance matrices generated by scheme A (left) and scheme B (right) with $N=20$ modes and loss $L_\mathrm{tot}=0.1$. We consider two random instances satisfying $S^*=1$ (top) and $S^* = -1$ (bottom), which are graphically depicted to the far left. In scheme A (B), we show the result after \num{4000} (\num{100}) round trips with $\eta=1.053$ ($\eta = 1.5$) and $T_\mathrm{A}=0.5$ ($T_\mathrm{B}=0.8$). For scheme A, the system diverges at $\eta=1.054$. The bottom-most plots show corresponding slices of the covariance matrix for mode \num{5} (i.e., $|\sigma^\alpha_{5,n}|$) for both the $S^*=1$ case (yellow circles) and the $S^*=-1$ case (blue triangles); the continuous lines are the results of fitting to the approximate model \eqref{model}.}
\label{fig:cmplot_circle}
\end{figure*}

In Fig.~\ref{fig:cmplot_circle}, we show the steady-state covariance matrices for both schemes A and B with typical system parameters. Using a set of randomly-generated couplings (with $S^*=\pm1$ as indicated), we apply transformations to the covariance matrix according to each element shown in Fig.~\ref{fig:schematics}; after a sufficient number of round trips, the system reaches steady state. From the figure, we see that the covariance matrices are well-approximated by \eqref{model}.

Note that $\sigma^\mathrm{x}$ for the frustrated case $S^*=-1$ has visibly smaller amplitude than the non-frustrated case $S^*=1$. Depending on whether there exists frustration, we can classify the couplings into 4 cases: (i) $S^*=1$ with even $N$, (ii) $S^*=-1$ with even $N$, (iii) $S^*=1$ with odd $N$, and (iv) $S^*=-1$ with odd $N$. It is generally known that entanglement is compromised by the existence of frustration~\cite{Wolf2004}. Also, systems with 1D ring geometry and an odd number of modes are known to exhibit less entanglement than the ones with even $N$~\cite{Adesso2006,Wolf2004}. We focus on the case (i) where there is no frustration in order to investigate the maximum capabilities of the schemes.

It is also worth mentioning that, so long as $S^*=1$, one can apply local phase flips $\hat{x}_k\mapsto S_{1,k}\hat{x}_k,~\hat{p}_k\mapsto S_{1,k}\hat{p}_k$ to make the couplings all ferromagnetic, (i.e., $s_k=1$ for all $k$), and we focus on this particular case without loss of generality. In this case, \eqref{model} becomes a translationally-invariant covariance matrix with exponentially-decaying correlations, and thus can be seen as a Gaussian matrix product state (GMPS)~\cite{Schuch2006,Schafer2012}.

From the form of the couplings, we expect a pair of EPR-type moments $\hat{u}^\mathrm{x}=N^{-1/2}\sum_{k=1}^N(-1)^k\hat{x}_k$ and $\hat{u}^\mathrm{p}=N^{-1/2}\sum_{k=1}^N\hat{p}_k$ to have the smallest variance (i.e., noise) in the $x$- and $p$-quadratures, respectively~\cite{Supplemental}. This situation is reminiscent of the condition for the Duan inequality in the context of bipartite systems~\cite{Duan2000}. In fact, by extending the work of Ref.~\cite{Maruo2016}, we can show that
\begin{align} \label{eq:entmeasure}
    K=2\sqrt{\langle(\hat{u}^\mathrm{x})^2\rangle\langle(\hat{u}^\mathrm{p})^2\rangle}\ge1
\end{align}
is a necessary condition for separability even for multimode systems like the CIM, and it turns out that if the covariance matrix takes the form \eqref{model} and the state is Gaussian, \eqref{eq:entmeasure} is also sufficient~\cite{Supplemental}. Since the systems we consider are well-approximated by \eqref{model}, we expect $K$ to be a particularly stringent measure we can use to quantify entanglement across the two schemes.

In a similar manner, we define $\hat{U}^\mathrm{x}=N^{-1/2}\sum_{k=1}^N\hat{x}_k$ and $\hat{U}^\mathrm{p}=N^{-1/2}\sum_{k=1}^N(-1)^k\hat{p}_k$, which we expect to have the largest variances~\cite{Supplemental}.

For scheme A, we can obtain analytic approximations for the variances of these moments if we make a simplifying assumption of translational invariance~\cite{Supplemental}. Denoting the variance of the moment $\hat u$ after $n$ round trips as $\langle \hat u^2 \rangle_n$, we find that (for $\alpha \in \{\text{x},\text{p}\}$)
\begin{subequations} \label{eq:schemea}
\begin{align}
    \langle (\hat{u}^\alpha{})^2\rangle_n &= w^\alpha+\left(\xi^\alpha\right)^n\left(\frac{1}{2}-w^\alpha\right) \\
    \langle (\hat{U}^\alpha{})^2\rangle_n &= W^\alpha+\left(\chi^\alpha\right)^n\left(\frac{1}{2}-W^\alpha\right),
\end{align}
\end{subequations}
where the geometric factors $0 < \xi^\alpha, \chi^\alpha \neq 1$ and constants $w^\alpha, W^\alpha \in \mathbb R$ can be analytically calculated~\cite{Supplemental}.

In particular, we find that $\chi^\text{x} = \eta^2(1-L_\text{tot})$, so for $\eta^2 > 1/\sqrt{1-L_\text{tot}}$, the system goes ``above threshold''; that is, the variance $\langle(\hat U^\text{x})^2\rangle_n$ becomes unbounded with $n$ (at least until saturation effects cause our Gaussian model to break down). Below threshold, the constants $w^\alpha$ and $W^\alpha$ can be interpreted as the steady-state variances $\langle(\hat u^\alpha)^2\rangle_{n\rightarrow\infty}$ and $\langle(\hat U^\alpha)^2\rangle_{n\rightarrow\infty}$, respectively.

\begin{figure}[h!b]
    \centering
    \includegraphics[width=0.44\textwidth]{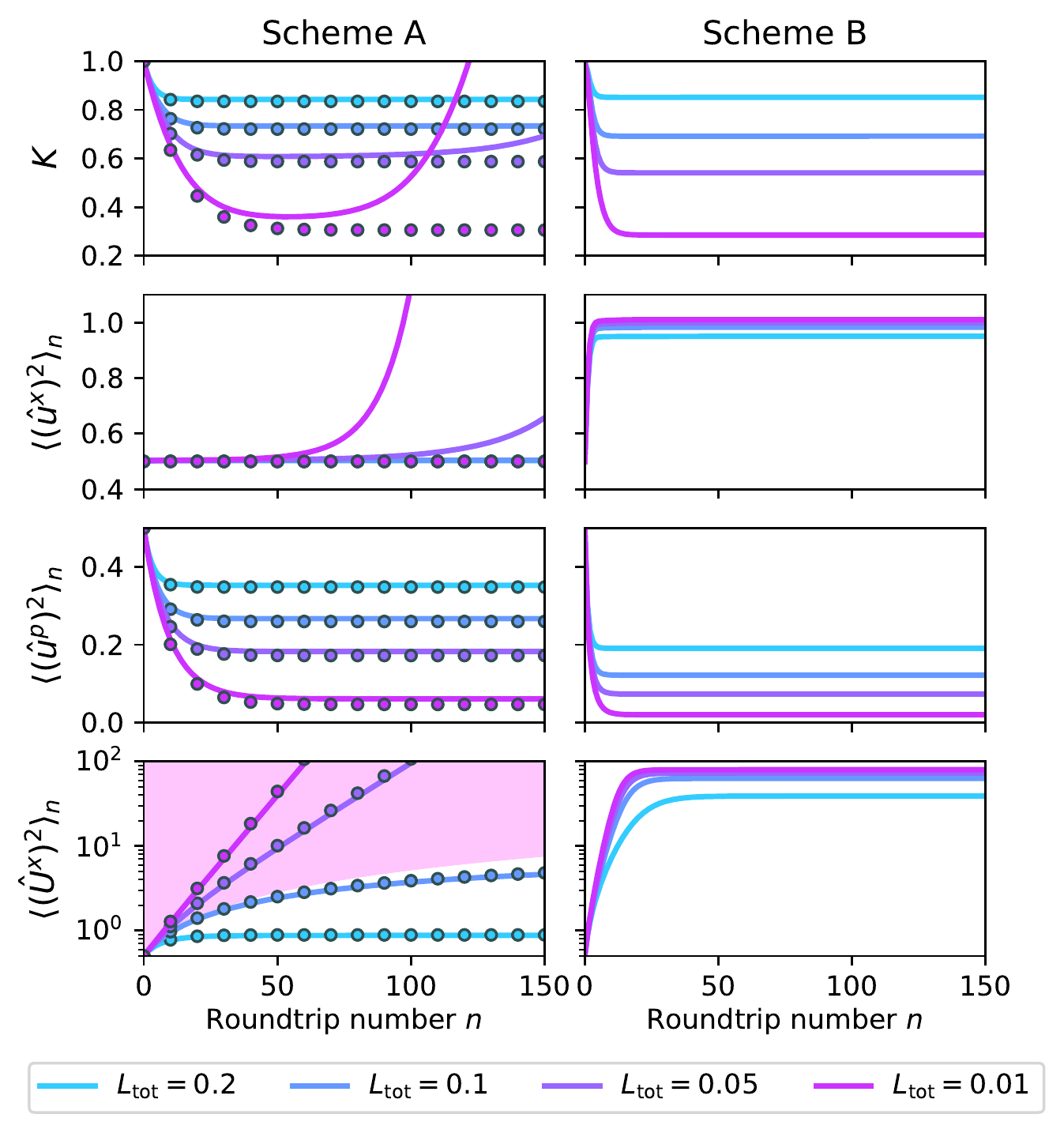}
    \caption{Time evolution of the entanglement measure $K$ and the characteristic variances $\langle (\hat{u}^\text{x})^2\rangle_n$, $\langle (\hat{u}^\text{p})^2\rangle_n$, and $\langle (\hat{U}^\text{x})^2\rangle_n$ for scheme A (left, using $\eta=1.05$ and $T_\mathrm{A}=0.5$) and scheme B (right, using $\eta=1.5$ and $T_\mathrm{B}=0.8$) for $N=30$ modes and various settings for the loss $L_\mathrm{tot}$. Solid lines denote simulations, while circles (scheme A only) are derived from the model \eqref{eq:schemea}. For scheme A, the system diverges for trajectories with $L_\mathrm{tot} < \num{0.093}$ (shaded region on bottom-left). Plot of $\langle (\hat{U}^\text{p})^2\rangle_n$ is omitted for brevity; see Ref.~\cite{Supplemental}.}
    \label{fig:transient}
\end{figure}

In Fig.~\ref{fig:transient}, we show how the variance of these moments evolve as a function of the number of round trips. For scheme A, the numerical results agree very well with the model \eqref{eq:schemea} below threshold, but above threshold, the model tends to deviate due to sensitivity of $\langle (\hat{u}^\text{x})^2\rangle$ to small non-translationally-invariant corrections. Furthermore, the numerical results indicate that when scheme A goes above threshold, $K$ eventually grows monotonically with time and thus exhibits entanglement only within a brief transient period. On the other hand, scheme B is more well-behaved, and it attains steady-state values $K < 1$ regardless of the amount of system loss.

\begin{figure}[b]
\centering
\includegraphics[width=0.45\textwidth]{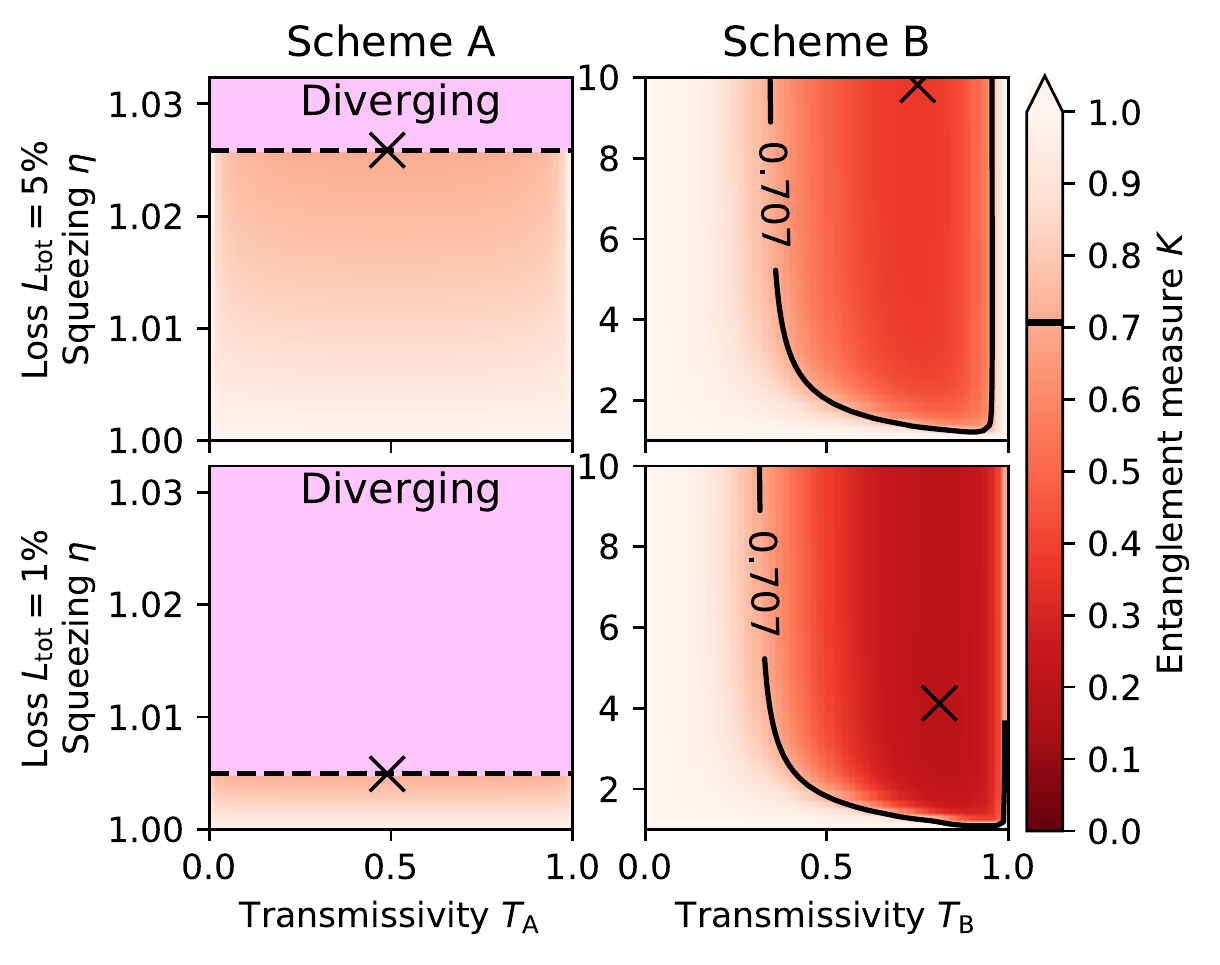}
\caption{Entanglement measure $K$ in schemes A (left) and B (right) as functions of the transmissivity $T_\mathrm{A,B}$ and squeezing $\eta$ for $N=30$ modes and loss $L_\mathrm{tot}=\SI{5}{\percent}$ (top) and $\SI{1}{\percent}$ (bottom). In the left plots, the shaded region bounded by the dashed line indicates where scheme A diverges. The solid line in the right plots show the minimal $K = K_\text{A0}$ attainable by scheme A. Black crosses mark the points with the smallest $K$. Note that the plots for scheme A are evaluated using \eqref{eq:schemea} since numerical simulations converge too slowly to be reliable due to a ``critical slowing down'' near threshold~\cite{Ballarini2009}; no calculations are done for the diverging region.}
\label{fig:fourplots}
\end{figure}

Using the analytic model \eqref{eq:schemea} for scheme A, the optimal steady-state value of $K$ that can be attained below threshold is $K_\mathrm{A0} = 1/\sqrt{2}=0.707\dots$, independently of the loss $L_\mathrm{tot}$~\cite{Supplemental}. This limiting value is attained for $T_\text{A} = 1/2$ and $\eta \rightarrow 1/\sqrt{1-L_\mathrm{tot}}$, and this point is shown graphically in parameter space to the left of Fig.~\ref{fig:fourplots}.

As shown to the right of Fig.~\ref{fig:fourplots}, however, scheme B supports steady states where $K$ is significantly smaller than $K_\mathrm{A0}$. Furthermore, whereas the valid (i.e., below-threshold) parameter range for scheme A is severely limited and hence requires precise control of $\eta$, scheme B shows only weak dependence on system parameters and is therefore more robust against variations.

\begin{figure}[h]
    \centering
    \includegraphics[width=0.49\textwidth]{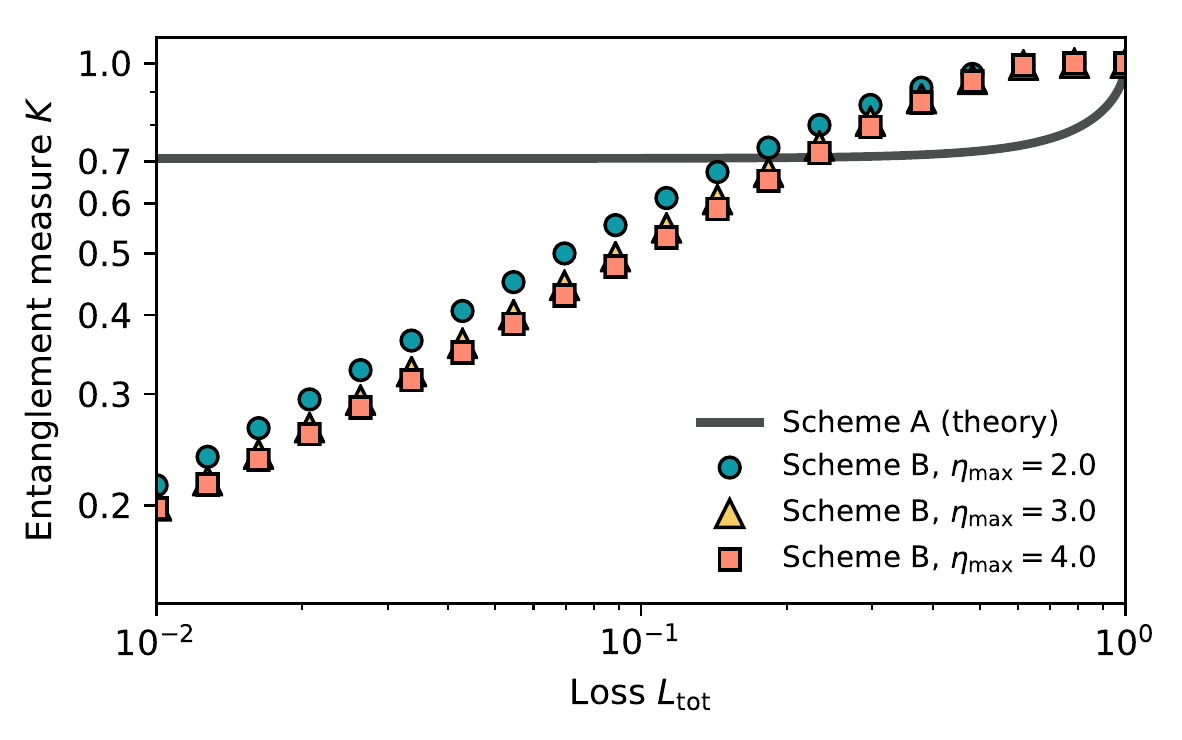}
    \caption{Best entanglement measure $K$ achievable as a function of loss $L_\mathrm{tot}$. Solid line: best achievable $K$ in scheme A using model \eqref{eq:schemea}~\cite{Supplemental}. Markers: $K_\mathrm{min}$ in scheme B based on numerical simulations with $N=30$ modes, assuming various limitations $\eta_\text{max}$ on the attainable squeezing.}
    \label{fig:kmin}
\end{figure}

From Fig.~\ref{fig:fourplots}, we also see that the optimal value of $\eta$ increases with increasing loss $L_\text{tot}$. However, in real experiments, technical limitations motivate the introduction of a bound $\eta_\mathrm{max}$ on the accessible squeezing parameter. Due to the insensitivity of $K$ to $\eta$ as shown in Fig.~\ref{fig:fourplots}, we expect this limitation to have mild consequences. We verify this in Fig.~\ref{fig:kmin}, which shows the best entanglement measure $K_\text{min}$ achievable in scheme B after optimizing over all $T_\text{B}$ and all $\eta \leq \eta_\text{max}$. Within the range of losses simulated, we observe scheme B shows monotonic linear decrease in $\log K_\mathrm{min}$ for decreasing $\log L_\mathrm{tot}$. In particular, for the no-loss case $L_\mathrm{tot}=0$ with $N=30$ modes, we get $K=1.4\times 10^{-2}$ with the optimal $\eta=1.15$ and $T_\mathrm{B}=99.4\%$. This can also be contrasted with a GMPS-generating scheme introduced in Ref.~\cite{Adesso2006} which requires large amounts of squeezing to reach this level of $K$~\cite{Supplemental}. 

Before concluding, we address some additional practical benefits of our measurement-based scheme. In scheme A, the coupling parities are controlled via the phase shift of the delay path, which in pulsed systems is usually implemented by a phase modulator; this usually increases the system loss. On the other hand, the coupling parities in scheme B are controlled via the choice of measurement quadratures, which can be implemented by phase-modulating the local oscillator and introduces no additional system loss. Furthermore, scheme B is amenable to integration with established solution-finding approaches based on measurement and feedback on the first-order moments $\langle\boldsymbol{\hat{x}}\rangle$~\cite{McMahon2016,Inagaki2016}. Because scheme B depends only on the covariance matrix, it can be made compatible with current CIM implementations by simply keeping track of the measurement backaction on $\langle\boldsymbol{\hat{x}}\rangle$~\cite{Supplemental}.

In this research, we have introduced a measurement-based scheme for generating entanglement inside a time-multiplexed CIM. We have shown this scheme to outperform conventional delay-line-based schemes in terms of the accessible entanglement, while also conferring additional practical benefits. Considering these benefits, this scheme motivates both theoretical and near-term experimental investigations into the computational advantages of CIMs with native multimode entanglement.

\section*{Acknowledgements}
This work is supported by the Army Research Office under Grant No.~W911NF-16-1-0086. PLM is supported by ImPACT Program of the Council for Science, Technology and Innovation (Cabinet Office, Government of Japan). RY, TO, and EN are supported by National Science Foundation under award PHY-1648807. RY is also supported by Masason Foundation. The authors would like to thank Logan G.\ Wright and Timoth\'{e}e Leleu for insightful discussions.

\pagebreak
\onecolumngrid
\begin{center}
\textbf{\large Supplementary Materials}
\end{center}
\setcounter{equation}{0}
\setcounter{figure}{0}
\setcounter{table}{0}
\setcounter{page}{1}
\setcounter{section}{0}
\makeatletter
\renewcommand{\theequation}{S\arabic{equation}}
\renewcommand{\thefigure}{S\arabic{figure}}
\renewcommand{\bibnumfmt}[1]{[S#1]}
\renewcommand{\citenumfont}[1]{S#1}

\section{Moments with smallest and largest variance}
\label{supermodes}
In this section, we show the derivation of the moments with smallest and largest variances $\hat{u}^\alpha$ and $\hat{U}^\alpha$ ($\alpha=\mathrm{x},\mathrm{p}$), respectively, assuming that the covariance matrix is well approximated by \eqref{model} in the main text. We assume an even $N$ number of modes with purely ferromagnetic couplings (i.e., $s_k=1$ for all $k$).

Because the covariance matrix $\sigma^\alpha$ has translational symmetry, its eigenvectors are the Bloch vectors
\begin{align}
    \boldsymbol{v}_j=\frac{1}{\sqrt{N}}\left(1,e^{2\pi\mathrm{i}j/N},\dots,e^{2\pi\mathrm{i}j(N-1)/N}\right)^\mathrm{T}
\end{align}
where $j=0,1,\dots,N-1$. The corresponding eigenvalues are
\begin{subequations}
\begin{align}
    v_j^\mathrm{x}=D^\mathrm{x}+\frac{d^\mathrm{x}\left\{1-(\gamma^\mathrm{x})^2\right\}\left\{1-(\gamma^\mathrm{x})^N\right\}}{1+(\gamma^\mathrm{x})^2-2\gamma^\mathrm{x}\cos (2\pi j/N)}
\end{align}
and
\begin{align}
    v_j^\mathrm{p}=D^\mathrm{p}-\frac{d^\mathrm{p}\left\{1-(\gamma^\mathrm{p})^2\right\}\left\{1-(\gamma^\mathrm{p})^N\right\}}{1+(\gamma^\mathrm{p})^2-2\gamma^\mathrm{p}\cos (2\pi j/N)}
\end{align}
\end{subequations}
for $\sigma^\mathrm{x}$ and $\sigma^\mathrm{p}$, respectively. Using the notation $\boldsymbol{\hat{x}}=(\hat{x}_1,\hat{x}_2,\dots,\hat{x}_N)^\mathrm{T}$ and $\boldsymbol{\hat{p}}=(\hat{p}_1,\hat{p}_2,\dots,\hat{p}_N)^\mathrm{T}$, the variance of a normalized zero-mean moment $\boldsymbol{v}^\mathrm{T}\boldsymbol{\hat{x}}$ is given by $\langle(\boldsymbol{v}^\mathrm{T}\boldsymbol{\hat{x}})^2\rangle=\boldsymbol{v}^\mathrm{T}\sigma^\mathrm{x}\boldsymbol{v}$ and similarly for $\boldsymbol{v}^\mathrm{T}\boldsymbol{\hat{p}}$.
Therefore, the moments with smallest variance for each quadrature are
\begin{align}
    \hat{u}^\mathrm{x}&=\boldsymbol{v}_{\frac{N}{2}}^\mathrm{T}\boldsymbol{\hat{x}}=\frac{1}{\sqrt{N}}\sum_{k=1}^{N}(-1)^k\hat{x}_k,&
    \hat{u}^\mathrm{p}&=\boldsymbol{v}_{0}^\mathrm{T}\boldsymbol{\hat{p}}=\frac{1}{\sqrt{N}}\sum_{k=1}^N\hat{p}_k,
\end{align}
while the ones with largest variance are
\begin{align}
    \hat{U}^\mathrm{x}&=\boldsymbol{v}_{0}^\mathrm{T}\boldsymbol{\hat{x}}=\frac{1}{\sqrt{N}}\sum_{k=1}^N\hat{x}_k,&\hat{U}^\mathrm{p}&=\boldsymbol{v}_{\frac{N}{2}}^\mathrm{T}\boldsymbol{\hat{p}}=\frac{1}{\sqrt{N}}\sum_{k=1}^N(-1)^k\hat{p}_k.
\end{align}

\section{Entanglement measure}
\label{inseparability}
Here, we derive and discuss the entanglement measure $K$ introduced in \eqref{eq:entmeasure} of the main text. Because the first-moments vector can be made zero by local operations, we assume $\langle\boldsymbol{\hat{R}}\rangle=0$ in the following. Let two normalized moments be
\begin{align}
    \hat{\mu}=\sum_{k=1}^N\mu_k\hat{x}_k,\quad\hat{\nu}=\sum_{k=1}^N\nu_k\hat{p}_k
\end{align}
where the coefficients $\mu_k$ and $\nu_k$ satisfy
\begin{align}
\label{conditions}
    \sum_{k=1}^N\mu_k^2=\sum_{k=1}^N\nu_k^2=1\quad\mathrm{and}\quad\mu_k^2=\nu_k^2.
\end{align}
We summarize these coefficients using the notations $\boldsymbol{\mu}=(\mu_1,\mu_2,\dots,\mu_N)^T$ and $\boldsymbol{\nu}=(\nu_1,\nu_2,\dots,\nu_N)^T$.

Now, if a state $\hat{\rho}$ is separable, it can be written as a weighted sum of separable density matrices such that
\begin{align}
    \hat{\rho}=\sum_m\lambda_m\hat{\rho}_{1}^{(m)}\otimes\hat{\rho}_{2}^{(m)}\dots\otimes\hat{\rho}_{N}^{(m)}
\end{align}
with $\sum_m\lambda_m=1$. Then, for this separable state $\hat{\rho}$, we calculate
\begin{align}
\begin{split}
J=&\langle\hat{\mu}^2\rangle+\langle\hat{\nu}^2\rangle\\
=&\sum_m\lambda_m(\langle\hat{\mu}^2\rangle_m+\langle\hat{\nu}^2\rangle_m)\\
=&\sum_m\sum_{k=1}^N\sum_{\ell=1}^N\lambda_m(\mu_k\mu_\ell\langle\hat{x}_k\hat{x}_\ell\rangle_m+\nu_k\nu_\ell\langle\hat{p}_k\hat{p}_\ell\rangle_m)
\end{split}
\end{align}
where the expectation $\langle~\cdot~\rangle_m$ is taken on the density matrix $\hat{\rho}^{(m)}=\hat{\rho}_{1}^{(m)}\otimes\hat{\rho}_{2}^{(m)}\otimes\dots\otimes\hat{\rho}_{N}^{(m)}$. We can rewrite this as
\begin{align}
J=&\sum_m\sum_{k=1}^N\sum_{\ell=1}^N\lambda_m\Bigl[\mu_k\mu_\ell\left(\langle\hat{x}_k\hat{x}_\ell\rangle_m-\langle\hat{x}_k\rangle_m\langle\hat{x}_\ell\rangle_m\right)+\nu_k\nu_\ell\left(\langle\hat{p}_k\hat{p}_\ell\rangle_m-\langle\hat{p}_k\rangle_m\langle\hat{p}_\ell\rangle_m\right)\Bigr]+W,
\end{align}
where
\begin{align}
\begin{split}
W&=\sum_m\sum_{k=1}^N\sum_{\ell=1}^N\lambda_m\Bigl(\mu_k\mu_\ell\langle\hat{x}_k\rangle_m\langle\hat{x}_\ell\rangle_m+\nu_k\nu_\ell\langle\hat{p}_k\rangle_m\langle\hat{p}_\ell\rangle_m\Bigr)\\
&=\sum_m\lambda_m\left[\left(\textstyle\sum_{k=1}^N\mu_k\langle\hat{x}_k\rangle_m\right)^2+\left(\textstyle\sum_{k=1}^N\nu_k\langle\hat{p}_k\rangle_m\right)^2\right]\ge0.
\end{split}
\end{align}
Since $\hat{\rho}^{(m)}$ is separable, we have $\langle\hat{x}_k\hat{x}_\ell\rangle_m=\langle\hat{x}_k\rangle_m\langle\hat{x}_\ell\rangle_m$ and $\langle\hat{p}_k\hat{p}_\ell\rangle_m=\langle\hat{p}_k\rangle_m\langle\hat{p}_\ell\rangle_m$ for any $\ell\ne k$. Thus the sum over $\ell$ collapses to the $\ell = k$ term, and we have
\begin{align}
    \begin{split}
J&=\sum_m\sum_{k=1}^N\lambda_m\Big[\mu_k^2\left(\langle\hat{x}_k^2\rangle_m-\langle\hat{x}_k\rangle_m^2\right)+\nu_k^2\left(\langle\hat{p}_k^2\rangle_m-\langle\hat{p}_k\rangle_m^2\right)\Big]+W\\
&=\sum_m\sum_{k=1}^N\lambda_m\mu_k^2\Big[\left(\langle\hat{x}_k^2\rangle_m-\langle\hat{x}_k\rangle_m^2\right)+\left(\langle\hat{p}_k^2\rangle_m-\langle\hat{p}_k\rangle_m^2\right)\Big]+W\\
&\ge\sum_m\lambda_m\sum_{k=1}^N\mu_k^2+W= 1+W\ge1.
\end{split}
\end{align}
The uncertainty principle $(\langle\hat{x}_k^2\rangle_m-\langle\hat{x}_k\rangle_m^2)+(\langle\hat{p}_k^2\rangle_m-\langle\hat{p}_k\rangle_m^2)\ge1$ is used to get the last line from the second line.
Thus, $J\ge1$ is a necessary condition for the separability of $\hat{\rho}$. This condition was first introduced in Ref.~\cite{S_Maruo2016}, and it can be seen as a multimode extension of the condition shown in Ref.~\cite{S_Duan2000}.

We can actually make the inequality more strict. Because local operations do not alter separability, we find that, more generally, $J(s) = s\langle\hat{\mu}^2\rangle+s^{-1}\langle\hat{\nu}^2\rangle\ge1$ has to be fulfilled for any $s$ if $\hat\rho$ is separable. (This expression corresponds to an arbitrary squeezing operation uniformly and locally applied to each mode.) By the Cauchy-Schwarz inequality, $J(s)$ attains its minimum for $s=\sqrt{\langle\hat{\nu}^2\rangle/\langle\hat{\mu}^2\rangle}$. Therefore, without loss of generality, a more stringent necessary condition for the separability of $\hat\rho$ is
\begin{align}
\label{kinequality}
    K=2\sqrt{\langle\hat{\mu}^2\rangle\langle\hat{\nu}^2\rangle}\ge1,
\end{align}
which is the entanglement measure for this work.

Generally, this condition becomes more stringent when we choose a pair of moments $\hat{\mu}$ and $\hat{\nu}$ with smaller variance. In particular, if we can take $\boldsymbol{\mu}$ and $\boldsymbol{\nu}$ as the eigenvectors of $\sigma^\mathrm{x}$ and $\sigma^\mathrm{p}$ with the smallest eigenvalues while fulfilling the condition \eqref{conditions} (which is not always possible), this latter choice is optimal. 

In fact, in this latter case, provided the two additional assumptions that 1) the state $\hat{\rho}$ is Gaussian, and 2) there are no correlation between the $x$- and $p$- quadratures, the inequality \eqref{kinequality} becomes both necessary and sufficient for the separability of $\hat{\rho}$. This is shown in the following way. Let us assume that \eqref{kinequality} holds for a choice of $\boldsymbol{\mu}$ and $\boldsymbol{\nu}$ such that they are the eigenvectors of $\sigma^\mathrm{x}$ and $\sigma^\mathrm{p}$ with the smallest eigenvalues. Then, we can apply appropriate uniform, local squeezing transformations so that the minimum eigenvalues of both $\boldsymbol{\mu}$ and $\boldsymbol{\nu}$ are at least $\frac{1}{2}$. Then if the $x$-$p$ correlations are zero, i.e., $\frac{1}{2}\langle\{\hat{x}_k,\hat{p}_\ell\}\rangle-\langle\hat{x}_k\rangle\langle\hat{p}_\ell\rangle=0$ for $1\le k,\ell\le N$, we have $(\sigma - 1/2) \ge 0$ for the entire covariance matrix (under that squeezing transformation). Because this latter result is a sufficient condition for the separability of a Gaussian state~\cite{S_Werner2001}, we have proven the assertion.

\section{Dynamics of the least and most noisy moments in scheme A}
\label{schemeAdynamics}
We consider the evolution of an even number $N$ optical modes under scheme A with squeezing parameter $\eta$, round trip loss $L_\mathrm{tot}$, and transmissivity $T_\mathrm{A}$ for the beamsplitters. Again, we assume that the couplings are all ferromagnetic. Although the exact solution is complicated (for example, it does not have an exact steady state because of its sequential nature), the dynamics can be well approximated by considering an effective system in which the operations occur simultaneously over each round trip. Inside the effective system, all the modes simultaneously experience the loss $L$ (where $L_\mathrm{tot}=1-(1-L)^2$) before and after the application of squeezing by $\eta$. Then, equal fractions of all the modes $k=1,2,\dots,N$ are extracted simultaneously by beamsplitters with transmissivity $T_\mathrm{A}$ and mixed with its subsequent mode $(k+1)$.

The simultaneous application of the loss before and after squeezing leads to a transformation of the covariance matrix according to
\begin{subequations}
\begin{align}
    \langle\hat{x}_k\hat{x}_\ell\rangle&\mapsto\eta^2(1-L)^2\langle\hat{x}_k\hat{x}_\ell\rangle+\frac{\delta_{k,\ell}}{2}(\eta^2L(1-L)+L) \\ \langle\hat{p}_k\hat{p}_\ell\rangle&\mapsto\eta^{-2}(1-L)^2\langle\hat{p}_k\hat{p}_\ell\rangle+\frac{\delta_{k,\ell}}{2}(\eta^{-2}L(1-L)+L).
\end{align}
\end{subequations}
Then, the simultaneous application of the beamsplitters leads to the inside-cavity modes
\begin{align}
    \hat{a}_k^\mathrm{i}{}=\sqrt{T_\mathrm{A}}\hat{a}_k+\sqrt{1-T_\mathrm{A}}\hat{a}_k^\mathrm{v},
\end{align}
and the outside-cavity modes
\begin{align}
    \hat{a}_k^\mathrm{o}{}=\sqrt{1-T_\mathrm{A}}\hat{a}_k-\sqrt{T_\mathrm{A}}\hat{a}_k^\mathrm{v},
\end{align}
where $\hat{a}_k^\mathrm{v}$ represents vacuum. Then, $\hat{a}_k^\mathrm{o}$ is mixed with $\hat{a}_{k+1}^\mathrm{i}$ to generate the intra-cavity modes
\begin{align}
    \hat{a}_k'=&(1-T_\mathrm{A})\hat{a}_k+T_\mathrm{A}\hat{a}_{k+1}+\sqrt{T_\mathrm{A}(1-T_\mathrm{A})}(\hat{a}_{k+1}^\mathrm{v}-\hat{a}_k^\mathrm{v}),
\end{align}
where the prime represents the modes after the mixture. We then repeat this procedure for the next roundtrip, using $\hat a'_k$ for the cavity modes.

We have empirically found that the covariance matrix produced by this set of maps is well approximated by $\eqref{model}$. This motivates us to consider the dynamics of the moments 
\begin{align}
    \hat{u}^\mathrm{x}&=\boldsymbol{v}_{\frac{N}{2}}^\mathrm{T}\boldsymbol{\hat{x}}=\frac{1}{\sqrt{N}}\sum_{k=1}^{N}(-1)^k\hat{x}_k,&
    \hat{u}^\mathrm{p}&=\boldsymbol{v}_{0}^\mathrm{T}\boldsymbol{\hat{p}}=\frac{1}{\sqrt{N}}\sum_{k=1}^N\hat{p}_k,
\end{align}
and
\begin{align}
    \hat{U}^\mathrm{x}&=\boldsymbol{v}_{0}^\mathrm{T}\boldsymbol{\hat{x}}=\frac{1}{\sqrt{N}}\sum_{k=1}^N\hat{x}_k,&\hat{U}^\mathrm{p}&=\boldsymbol{v}_{\frac{N}{2}}^\mathrm{T}\boldsymbol{\hat{p}}=\frac{1}{\sqrt{N}}\sum_{k=1}^N(-1)^k\hat{p}_k.
\end{align}
The evolution of these moments are given by
\begin{align}
    \hat{u}^\mathrm{x}{}'&=(1-2T_\mathrm{A})\hat{u}^\mathrm{x}-2\sqrt{T_\mathrm{A}(1-T_\mathrm{A})/N}\sum^N_{k=1}(-1)^k\hat{x}^\mathrm{v}_k,&\hat{u}^p{}'&=\hat{u}^\mathrm{p},\\
    \hat{U}^\mathrm{x}{}'&=\hat{U}^x,&\hat{U}^\mathrm{p}{}'&=(1-2T_\mathrm{A})\hat{U}^\mathrm{p}-2\sqrt{T_\mathrm{A}(1-T_\mathrm{A})/N}\sum^N_{k=1}(-1)^k\hat{p}^\mathrm{v}_k.
\end{align}
As a result, we get the recursive equations
\begin{subequations}
\begin{align}
    \langle (\hat{u}^\mathrm{x})^2\rangle_{n+1}&=\eta^2(1-L)^2(2T_\mathrm{A}-1)^2\langle (\hat{u}^\mathrm{x})^2\rangle_{n}+\frac{1}{2}(2T_\mathrm{A}-1)^2(\eta^2L(1-L)+L)+2T_\mathrm{A}(1-T_\mathrm{A}),\\
    \langle (\hat{u}^\mathrm{p})^2\rangle_{n+1}&=\eta^{-2}(1-L)^2\langle (\hat{u}^\mathrm{p})^2\rangle_{n}+\frac{1}{2}(\eta^{-2}L(1-L)+L),\\
    \langle (\hat{U}^\mathrm{x})^2\rangle_{n+1}&=\eta^2(1-L)^2\langle (\hat{U}^\mathrm{x})^2\rangle_{n}+\frac{1}{2}(\eta^2L(1-L)+L),\\
    \langle (\hat{U}^\mathrm{p})^2\rangle_{n+1}&=\eta^{-2}(1-L)^2(2T_\mathrm{A}-1)^2\langle (\hat{U}^\mathrm{p})^2\rangle_{n}+\frac{1}{2}(2T_\mathrm{A}-1)^2(\eta^{-2}L(1-L)+L)+2T_\mathrm{A}(1-T_\mathrm{A}),
\end{align}
\end{subequations}
where $n$ denotes the round-trip number. These equations can be exactly solved to produce
\begin{subequations}
\begin{align}
\langle (\hat{u}^\alpha)^2\rangle_n&=w^\mathrm{\alpha}+(\xi^\alpha)^n\left(\frac{1}{2}-w^\mathrm{\alpha}\right)\\
\langle (\hat{U}^\alpha)^2\rangle_n&=W^\alpha+(\chi^\alpha)^n\left(\frac{1}{2}-W^\alpha\right),
\end{align}
\end{subequations}
with the geometric factors
\begin{subequations}
\begin{align}
    &\xi^\mathrm{x}=\eta^{2}(1-L)^2(2T_\mathrm{A}-1)^2,&\xi^\mathrm{p}&=\eta^{-2}(1-L)^2,\\
    &\chi^\mathrm{x}=\eta^2(1-L)^2,&\chi^\mathrm{p}&=\eta^{-2}(1-L)^2(2T_\mathrm{A}-1)^2,
\end{align}
\end{subequations}
and
\begin{subequations}
\begin{small}
\begin{align}
    w^\mathrm{x}&=\frac{4T_\mathrm{A}(1-T_\mathrm{A})+(2T_\mathrm{A}-1)^2\left\{\eta^2L(1-L)+L\right\}}{2(1-\xi^\mathrm{x})},&w^\mathrm{p}&=\frac{\eta^{-2}L(1-L)+L}{2(1-\xi^\mathrm{p})},\\
W^\mathrm{x}&=\frac{\eta^2L(1-L)+L}{2(1-\chi^\mathrm{x})},&W^\mathrm{p}&=\frac{4T_\mathrm{A}(1-T_\mathrm{A})+(2T_\mathrm{A}-1)^2\left\{\eta^{-2}L(1-L)+L\right\}}{2(1-\chi^\mathrm{p})}
\end{align}
\end{small}\end{subequations}as long as both $\xi^\mathrm{x}\ne1$ and $\chi^\mathrm{x}\ne1$. For the two special cases $\xi^\mathrm{x}=1$ or $\chi^\mathrm{x}=1$, $\langle (\hat{u}^\mathrm{x})^2\rangle_n$ and $\langle (\hat{U}^\mathrm{x})^2\rangle_n$ depend linearly on $n$ respectively as
\begin{subequations}
\begin{align}
    &\langle (\hat{u}^\mathrm{x})^2\rangle_n=\frac{1}{2}+n\left[\frac{1}{2}(2T_\mathrm{A}-1)^2(\eta^2L(1-L)+L)+2T_\mathrm{A}(1-T_\mathrm{A})\right]~&\mathrm{for}&~\xi^\mathrm{x}=1\\
    &\langle (\hat{U}^\mathrm{x})^2\rangle_n=\frac{1}{2}+\frac{n}{2}(\eta^2L(1-L)+L)~&\mathrm{for}&~\chi^\mathrm{x}=1.
\end{align}
\end{subequations}
As shown in the main text, this effective model agrees with numerical results as long as the system does not diverge.

Because $\chi^\mathrm{x}<1$ is the necessary and sufficient condition for the existence of the steady state in this model, $\eta=1/(1-L)=1/\sqrt{1-L_\mathrm{tot}}$ is the divergence threshold. Assuming that the system is not diverging ($\eta<1/\sqrt{1-L_\mathrm{tot}}$), the entanglement measure $K$ of a steady state of scheme A with loss $L$ is bounded from below as
\begin{small}
\begin{align}
    K=2\sqrt{\frac{\left[4T_\mathrm{A}(1-T_\mathrm{A})+(2T_\mathrm{A}-1)^2\left\{\eta^2L(1-L)+L\right\}\right][\eta^{-2}L(1-L)+L]}{4\left\{1-\eta^2(1-L)^2(2T_\mathrm{A}-1)^2\right\}\left\{1-\eta^{-2}(1-L)^2\right\}}}\ge\sqrt{\frac{\eta^{-2}L(1-L)+L}{1-\eta^{-2}(1-L)^2}}>\sqrt{1-\frac{\sqrt{1-L_\mathrm{tot}}}{2-L_\mathrm{tot}}}.
\end{align}
\end{small}
The first inequality is attained for $T_\mathrm{A}=1/2$ while the second inequality saturates at $\eta\rightarrow1/\sqrt{1-L_\mathrm{tot}}$.
\section{Comparison with GMPS generation protocols}
\begin{figure*}[ht]
\centering
\includegraphics[width=1.0\textwidth]{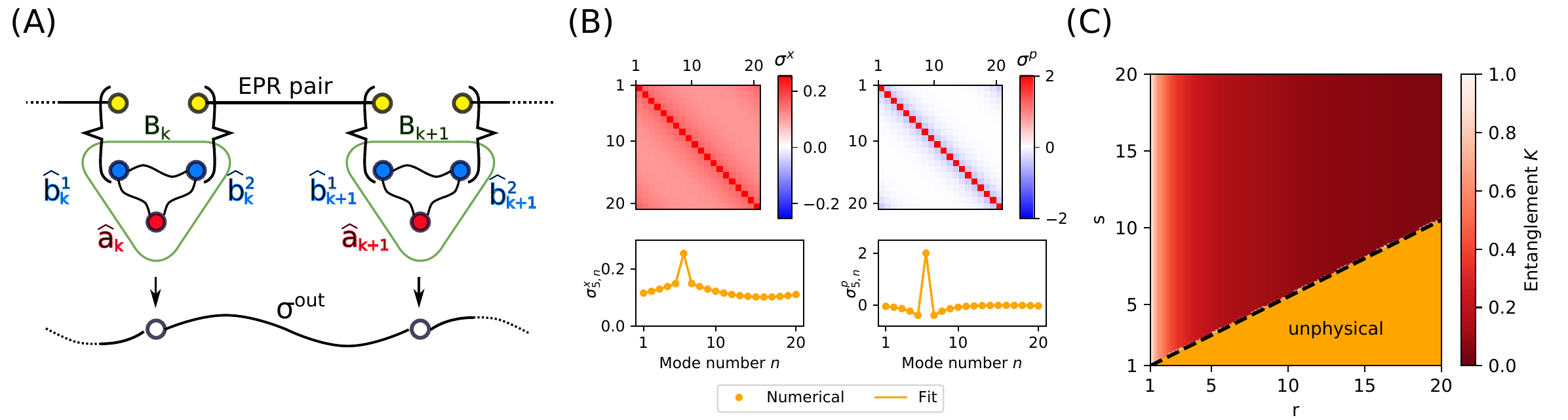}
\caption{(A) Schematic for the construction of a GMPS covariance matrix $\sigma^\mathrm{out}$. Building blocks $B_k$ and $B_{k+1}$ are entangled by projecting the modes $\hat{b}_{k+1}^\mathrm{1}$ and $\hat{b}_k^\mathrm{2}$ onto infinitely squeezed EPR-pairs. Figure adapted from Ref.~\cite{S_Adesso2006}. (B) Covariance matrix of a numerically-generated GMPS using $s=10$ and $r=5$ with $N=20$ modes. (C) Numerically calculated entanglement measure $K$ for $N=30$ as a function of parameters $r$ and $s$. The orange region represents unphysical combinations of the parameters.}
\label{fig:gmps_combined}
\end{figure*}

Our scheme is closely related to the construction of Gaussian matrix product states (GMPS) presented in Ref.~\cite{S_Adesso2006}. We briefly review the GMPS construction here, which we denote as scheme C. In scheme C, we can generate an $N$-mode ($N$ assumed even) GMPS using $N$ copies of three-mode building blocks $B_k~(k=1,2,\dots,N)$ as shown in Fig.~\ref{fig:gmps_combined}(A). Each building block $B_k$ is composed of two auxiliary modes $\hat{b}_k^\mathrm{1}$ and $\hat{b}_k^\mathrm{2}$ and an output mode $\hat{a}_k$. The $\alpha$-quadrature ($\alpha=\mathrm{x},\mathrm{p}$) covariance matrix $\sigma^{\mathrm{B}_k,\alpha}$ of the building block $B_k$ has the form
\begin{subequations}
\begin{align}
    &\sigma^{\mathrm{B}_k,\mathrm{x}}=\frac{1}{2}
    \begin{pmatrix}s&t_+&u_+\\
        t_+&s&u_+\\
        u_+&u_+&r
    \end{pmatrix},&\sigma^{\mathrm{B}_k,\mathrm{p}}=\frac{1}{2}
    \begin{pmatrix}s&t_-&u_-\\
        t_-&s&u_-\\
        u_-&u_-&r
    \end{pmatrix}
\end{align}
where
\begin{align}
    t_\pm&=\frac{1}{4s}\left(r^2-1\pm\sqrt{16s^4-8s^2(1+r^2)+(r^2-1)^2}\right) \\
    u_\pm&=\frac{1}{4}\sqrt{\frac{r^2-1}{sr}}\left(\sqrt{(r-2s)^2-1}\pm\sqrt{(r+2s)^2-1}\right).
\end{align}
\end{subequations}
Here, the modes are ordered as $\hat{b}_k^\mathrm{1}$, $\hat{b}_k^\mathrm{2}$, and $\hat{a}_k$. Note that $r\ge1$ and $s\ge(r+1)/2$ have to be fulfilled for the building blocks to be physical, but they are otherwise not constrained. In the construction, the output mode $\hat a_k$ is entangled with the output mode $\hat a_{k+1}$ by projecting the auxiliary modes $\hat{b}^\mathrm{1}_{k+1}$ and $\hat{b}^\mathrm{2}_{k}$ onto an infinitely-squeezed EPR pair, as depicted by the two yellow circles connected by a line in Fig.~\ref{fig:gmps_combined}(A). (The correlation of this entanglement is determined by the correlation of the EPR pair.) After these projections, we have an output state consisting of the output modes $\hat{a}_k$, which is identified as a GMPS covariance matrix $\sigma^\mathrm{out}$. In Fig.~\ref{fig:gmps_combined}(B), we show such a covariance matrix generated under scheme C with typical parameters, and it can be seen that this covariance matrix is well approximated by the model \eqref{model} in the main text. Note that all of the process assumes no loss, i.e., $L_\mathrm{tot}=0$.

Here, we can find the moments with smallest variance $\hat{u}^\alpha$ and ones with largest variance $\hat{U}^\alpha$ in the same way as in schemes A and B. According to Ref.~\cite{S_Adesso2006}, the GMPS covariance matrix can be written as
\begin{align}
    \sigma^\text{out} = \sigma^\text{a} - (\sigma^\text{ab})^\mathrm{T} \left(\sigma^\text{b} + \theta\sigma^\text{EPR}\theta\right)^{-1}\sigma^\text{ab},
\end{align}
where $\sigma^\text{a}$ is the covariance matrix of the output modes $\hat a_k$ and $\sigma^\text{b}$ is the covariance matrix of the auxiliary modes $\hat b^1_k$ and $\hat b^2_k$, such that $\bigoplus_{k=1}^N \sigma^{B_k} = \begin{pmatrix}\sigma^\text{b}&\sigma^\text{ab} \\ (\sigma^\text{ab})^\mathrm{T}&\sigma^\text{a}\end{pmatrix}$, while $\sigma^\text{EPR}$ is the covariance matrix of the EPR pairs. Here, $\theta = \bigoplus_{k=1}^N \diag(1,-1)$. Using this result, we can thus calculate $\langle(\hat u^\alpha)^2\rangle = (\boldsymbol{u}^\alpha)^\mathrm{T} \sigma^\text{out}\boldsymbol{u}^\alpha$ (and similarly for $\langle(\hat U^\alpha)^2\rangle$).

Interestingly, though $\sigma^\mathrm{out}$ can take a complicated form, the analytic expressions for the uncertainties of the characteristic moments can be shown to be
\begin{align}
    \langle (\hat{u}^\mathrm{x})^2\rangle=\langle(\hat{u}^\mathrm{p})^2\rangle=\frac{1}{2r},\quad\langle(\hat{U}^\mathrm{x})^2\rangle=\langle(\hat{U}^\mathrm{p})^2\rangle=\frac{r}{2},
\end{align}
without approximation. As a result, for scheme C, we have $K=1/r$. This is supported by the numerical result shown in Fig.~\ref{fig:gmps_combined}(C). This result also indicates that the state is always entangled when $r>1$, which is consistent with the result found  in Ref.~\cite{S_Adesso2006} based on the ``positivity of partial transposition" criteria. Finally, this scaling of $K=1/r$ means that one needs to have infinitely large squeezing in the original state in order to get $K\rightarrow0$; this is in a contrast to scheme B where $K$ is minimized at a finite squeezing $\eta$.

\section{Accessibility of mode amplitudes in Scheme B}
\label{appendix:problemsolving}
The solution-finding procedure used in CIMs that have experimentally been realized at a large scale relies on the measurement of and the feedback on the first-order moments $\langle\boldsymbol{\hat{x}}\rangle$~\cite{S_McMahon2016,S_Inagaki2016}. On the other hand, in scheme B, the entanglement generation setup produces measurements of $\hat{y}_k=(\hat{x}_k-s_k\hat{x}_{k+1})/\sqrt{2}$, up to constant factors and vacuum noise. In principle, we can simply perform the standard solution-finding procedure using another outcoupling port to measure $\langle\boldsymbol{\hat{x}}\rangle$, but if there is an invertible transformation $M$ connecting the measured basis $\boldsymbol{\hat{y}}$ and the original basis $\boldsymbol{\hat{x}}$ such that $\boldsymbol{\hat{y}}=M\boldsymbol{\hat{x}}$, we can also reuse the results of measuring $\hat y_k$ with scheme B to infer $\boldsymbol{\hat{x}}$.

The form of the transformation matrix is
\begin{align}
    M=\frac{1}{\sqrt{2}}\begin{pmatrix}
    1&-s_1&0&\cdots&0\\
    0&1&-s_2&\cdots&0\\
    \vdots&\vdots&\vdots&\ddots&\vdots\\
    -s_N&0&0&\cdots&1
    \end{pmatrix},
\end{align}
which can be shown to be invertible if and only if $S^*=s_1s_2\dots s_N=-1$, upon which we have
\begin{align}
\begin{small}
    M^{-1}=\frac{1}{\sqrt{2}}\begin{pmatrix}
    1&S_{1,2}&S_{1,3}&\cdots&S_{1,N}\\
    -S_{1,2}&1&S_{2,3}&\cdots&S_{2,N}\\
    -S_{1,3}&-S_{2,3}&1&\cdots&S_{3,N}\\
    \vdots&\vdots&\vdots&\ddots&\vdots\\
    -S_{1,N}&-S_{2,N}&-S_{3,N}&\cdots&1
    \end{pmatrix}
    \end{small},
\end{align}
where $S_{k,\ell}=\prod_{i=\min(k,\ell)}^{\max{(k,\ell)}-1}s_i~(1\le k,\ell\le N)$. If this condition is met, then the first-order moments are given by
\begin{align}
\langle\boldsymbol{\hat{x}}\rangle =M^{-1}\langle\boldsymbol{\hat{y}}\rangle.
\end{align}

\end{document}